\documentclass[final,3p,times]{elsarticle}

\usepackage{lineno}

\usepackage{amssymb}
\usepackage{amsthm}
\usepackage{amsmath}
\usepackage{multicol}
\usepackage{multirow}
\usepackage{float}
\usepackage{hyperref}
\usepackage{fancyref}
\usepackage{graphicx}
\usepackage{subcaption}
\usepackage{geometry}

\begin{document}

\begin{frontmatter}


\title{The study of gamma-radiation induced displacement damage in $n^+$-in-$p$ silicon diodes}


\author[ASCR]{M. Mikeštíková\corref{cor}}
\cortext[cor]{Corresponding author}
\ead{mikestik@fzu.cz}

\author[ASCR]{P. Federičová}
\author[UJP]{P. Gallus}
\author[ASCR]{\fnref{MFF}R. Jirásek}
\author[ASCR]{J. Kozáková}
\author[ASCR]{J. Kroll}
\author[ASCR]{J. Kvasnička}
\author[ASCR]{\fnref{MFF}V. Latoňová}
\author[JSI]{I. Mandić}
\author[MFF]{K.~Mašek}
\author[ASCR]{\fnref{ISR}P.~Novotný}
\author[ASCR]{\fnref{UPOL}R. Přívara}
\author[ASCR]{P. Tůma}
\author[ASCR]{\fnref{FRB}I. Zatočilová}


\affiliation[ASCR]{organization={Institute of Physics, Academy of Sciences of the Czech Republic},
            addressline={Na Slovance 1999/2}, 
            city={Prague 8},
            postcode={18200},
            country={Czech Republic}}
\affiliation[UJP]{organization={UJP Praha a.s.},
            addressline={Nad Kamínkou 1345}, 
            city={Prague 5 - Zbraslav},
            postcode={15610},
            country={Czech Republic}}
\affiliation[MFF]{organization={Faculty of Mathematics and Physics, Charles University},
            addressline={V Holešovičkách 742/2}, 
            city={Prague 8},
            postcode={18000},
            country={Czech Republic}}
\affiliation[JSI]{organization={Experimental Particle Physics Department, Jožef Stefan Institute},
            addressline={Jamova cesta 39}, 
            city={Ljubljana},
            postcode={SI-1000},
            country={Slovenia}}

\affiliation[ISR]{organization={Faculty of Physics, Weizmann Institute of Science},
            addressline={234 Herzl Street}, 
            city={Rehovot},
            postcode={7610001},
            country={Israel}}

\affiliation[UPOL]{organization={Joint Laboratory of Optics of Palacky University and Institute of Physics of the Czech Academy of Sciences},
            addressline={17. listopadu 1154/50a}, 
            city={Olomouc},
            postcode={77900},
            country={Czech R.}}
            
\affiliation[FRB]{organization={now at Physikalisches Institut, Albert-Ludwigs-Universität Freiburg},
            addressline={Hermann-Herder-Straße 3}, 
            city={Freiburg},
            postcode={79104},
            country={Germany}}

\begin{abstract}

The bulk damage of $p$-type silicon sensors caused by gamma irradiation with high total ionizing doses has been investigated. The~study was carried out on different types of $n^+$-in-$p$ silicon diodes with different oxygen concentrations and silicon bulk resistivities.  The Secondary-Ion Mass Spectrometry technique was used to determine the relative concentration of oxygen in the individual samples. The measured diodes were irradiated by a $^{60}$Co gamma source to total ionizing doses ranging from 0.50 up to 8.28 MGy, and annealed for 80 minutes at 60~°C. The main goal of the study was to characterize the gamma-radiation induced displacement damage by measuring \textit{I--V} and \textit{C--V} characteristics, and the evolution of the full depletion voltage with the total ionizing dose. The Transient Current Technique was used to verify the full depletion voltage and to extract the electric field distribution and the sign of the space charge in the silicon diodes irradiated to the lowest and the highest delivered total ionizing doses. 

The results show a linear increase of the bulk leakage current with the total ionizing dose, with the damage coefficient being dependent on initial resistivity and oxygen concentration of the silicon diode. The effective doping concentration and full depletion voltage decrease significantly with an increasing total ionizing dose, before starting to increase again at a specific dose. We assume that the initial decrease in the effective doping concentration is caused by the~effect of acceptor removal. An additional notable finding of~this study is that the bulk leakage current and \textit{C--V} characteristics of the gamma-irradiated diodes do not show any evidence of an~annealing effect.
\end{abstract}
\end{frontmatter}


\section*{Introduction} \label{sec:intro}

Silicon detectors employed for the inner tracking region in forthcoming large particle accelerator experiments will be subjected to a demanding radiation environment. The high particle fluences and ionizing doses trigger electrically active defects in the silicon bulk, as well as SiO$_2$ insulating layer, and in Si-SiO$_2$ interface, ultimately deteriorating the sensor surface. New radiation hard technology was developed using boron doped $p$-type silicon bulk with an $n^+$ implanted surface ($n^+$-in-$p$).

The primary mechanism of bulk damage in silicon detectors, induced by hadrons, electrons, or higher-energy gamma rays, involves displacing the Primary Knock-on Atom (PKA) from its lattice site. The minimum transfer energy required for this process is $\approx$ 25~eV. The minimum energy required to generate such a single displacement resulting in a silicon interstitial-vacancy pair (Frenkel pair) for particles like neutrons and protons  is $\sim$ 175~eV and for electrons $\sim$~260~keV. For our investigation, silicon samples were irradiated by $^{60}$Co~$\gamma$-rays. In this case, the displacement damage is caused primarily by the interaction of Compton electrons with energies of a few hundred of~keV (up~to $\approx$~1~MeV) with the silicon lattice. Production of clusters of defects is not possible as the minimum electron energy needed for the production of clusters is $\approx$~8~MeV \cite{bib:hartmann} and the maximum recoil energy for the Si PKA by the Compton electron is only around 140~eV. Therefore, the damage observed in our silicon samples is solely attributed to the point defects.
Radiation effects after hadron irradiations on space charge, leakage current, and other characteristics of $p$-type materials with different oxygen concentration, resistivity, and doping have been investigated for a long time \cite{bib:pirollo, bib:casse1, bib:segneri}.

The $^{60}$Co~$\gamma$-ray induced displacement damage of silicon detectors was previously studied in~detail for $n$-type bulk \cite{bib:lindstrom, bib:pintilie}. Considering the widespread utilization of $p$-type bulk silicon technology for experiments at the High Luminosity LHC, there is a~need for a comprehensive study on the gamma induced damage of $p$-type silicon devices.  Initial studies on $p$-type silicon samples irradiated with gamma rays up to 3~MGy indicated a~noticeable decrease of the full depletion voltage ($V_{\mathrm{FD}}$) and effective doping concentration ($N_{\mathrm{eff}}$) with an increasing total ionizing dose (TID) \cite{bib:mikestikova}. In our previous study \cite{bib:zatocilova} the bulk radiation damage of the high resistivity $p$-type silicon samples was investigated up to an extended TID of 8.3~MGy delivered by $^{60}$Co~$\gamma$-rays. The main goal of the presented research was to extend this study of the radiation-induced displacement damage by a~detailed interpretation of measured \textit{I--V} and \textit{C--V} characteristics, with a special focus put on~the evolution of $V_{\mathrm{FD}}$ with TID, and a determination of the relation between 1~MeV neutron equivalent fluence and TID delivered by gamma irradiation.  Additionally, the~Transient Current Technique method (TCT) was used to extract the electric field distribution and verify the estimation of $V_{\mathrm{FD}}$. 


\section{Samples and irradiations} \label{sec:samples}
The study was carried out on three types of $n^+$-in-$p$ standard float zone diodes (labeled A, B, and C) provided by three different manufacturers. These diodes have similar active areas ranging from 0.0142 to 0.0149 cm$^3$ and thicknesses between 285 and 290 $\mu$m, but they vary in silicon bulk resistivity. The~initial full depletion voltage was estimated by conducting \textit{C--V} measurements on several unirradiated diodes of each type (refer to 2.2). The estimated $V_{\mathrm{FD}}$ is 283.6$\pm$12.0 V for diode A, 273.4$\pm$10.7 V for B, and 36.9$\pm$8.3 V for diode C. The initial silicon bulk resistivity $\rho$ was determined from $V_{\mathrm{FD}}$ using the~equation
\begin{equation*}
    \rho = \frac{d^2}{2 \varepsilon \mu V_{FD}},
\end{equation*}

\noindent where $\mu$ is the hole mobility, $\varepsilon$  the permittivity of silicon, and $d$ represents the active thickness of the diode. Diodes A and B exhibited similar $V_{\mathrm{FD}}$, and thus also comparable resistivity values of 3.1$\pm$0.1 and 3.3$\pm$0.1 k$\Omega\cdot$cm, respectively. In contrast, sample C had significantly higher initial resistivity of 24.0$\pm$4.0 k$\Omega\cdot$cm, which means lower boron doping, or significant compensation of $p$-type bulk by e.g. phosphorus. According to the manufacturer’s specifications, the~wafer oxygen concentration for samples A and B ranges from 1.5 $\times$ 10$^{16}$ to 6.5 $\times$ 10$^{17}$ atoms/cm$^3$. Unfortunately, the oxygen concentration for sample C was not provided. 

The Secondary Ion Mass Spectroscopy (SIMS) was utilized to determine the relative concentration of oxygen in the individual samples to a maximal depth of 14 $\mu$m. Cesium ions with an energy of 7~keV were employed as the primary ion source. The results show that the concentration of oxygen decreases with an increasing depth of diode for all three samples. The decrease in oxygen concentration is least pronounced in sample C and most significant in sample A. At a depth of 14 $\mu$m, sample C has the highest oxygen concentration, while sample A has the lowest. 

The diodes were irradiated by a $^{60}$Co gamma source at UJP PRAHA a.s. \cite{bib:ujp} up to the maximum TID of 8.28 MGy. During the irradiation, the diodes were placed in the charge particle equilibrium (CPE) box in accordance with the recommendation of ESA/SCC \cite{bib:esa}. This way the dose enhancement from low-energy scattered radiation is minimized by producing electron equilibrium and a uniform distribution of energy deposited in the irradiated diodes is ensured. The dose rate (in silicon) ranged from 160 to 190~Gy/min in the individual irradiation campaigns, with an estimated uncertainty of less than 5 \%. This in-silicon dose rate was calculated from the in-air dose rate measured by a calibrated ionizing chamber under the same conditions as the tested samples. Cooling with an air fan kept the temperature below 35~°C throughout the irradiation. After the irradiation, the samples were promptly refrigerated at temperatures below --20~°C to prevent uncontrolled annealing.

\section{Experimental methods and results} \label{sec:met-res}

\begin{figure}[h]
\centering
\begin{subfigure}{0.7\textwidth}
\centering
\includegraphics[width=10cm]{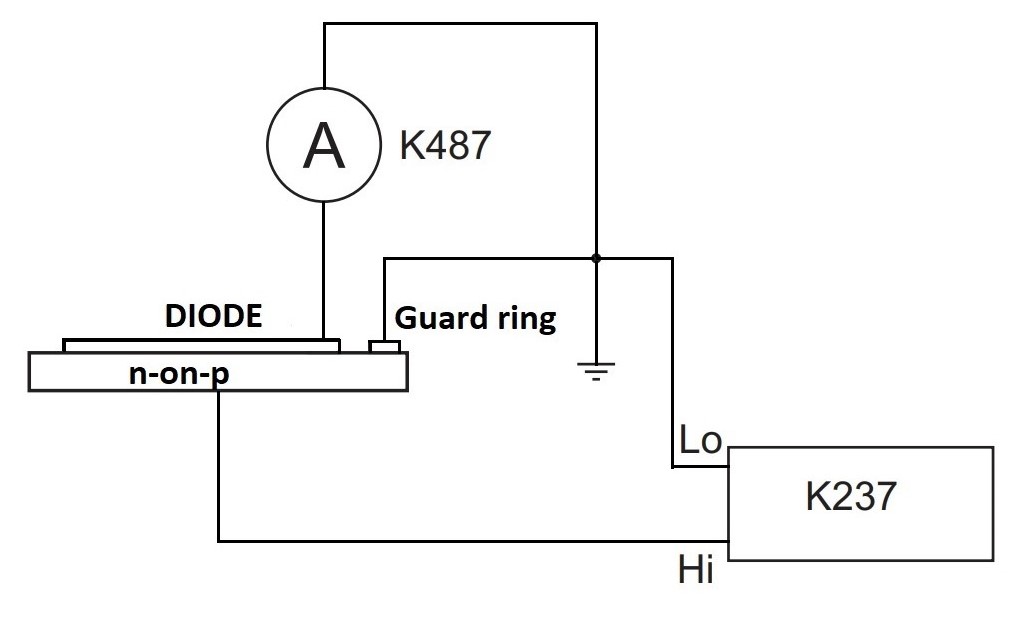} 
\caption{}
\label{fig:IV-setup}
\end{subfigure}
\begin{subfigure}{0.7\textwidth}
\centering
\includegraphics[width=10cm]{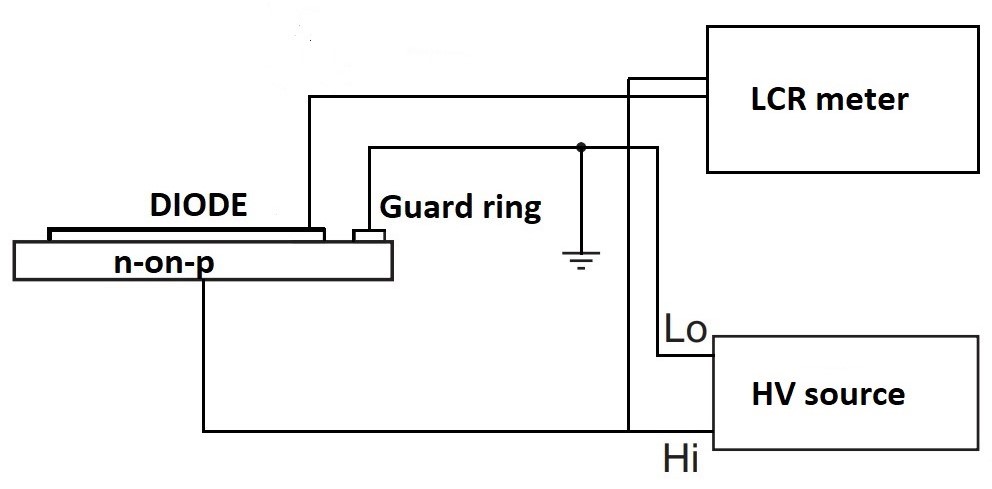}
\caption{}
\label{fig:CV-setup}
\end{subfigure}
\caption{Experimental setup used for measurement of \textit{I--V} (a) and \textit{C--V} (b) characteristics of the studied silicon diodes.}
\label{fig:setup}
\end{figure}

To evaluate the effects of bulk radiation damage induced by $^{60}$Co~$\gamma$-rays, the \textit{I--V} and \textit{C--V} characteristics of the diodes were measured before and after their irradiation. Great care was taken to properly determine the bulk leakage current flowing exclusively through the active volume of the diode bounded by the guard ring, by subtracting parasitic surface currents dominating outside of this guard ring. Fig. \ref{fig:IV-setup} schematically shows the experimental setup used for the measurement of \textit{I--V} characteristics, which provides the value of total current $I_{\mathrm{tot}}$, as well as the value of bulk leakage current $I_{\mathrm{bulk}}$. The total current is defined as a sum of bulk leakage current and surface current.
\begin{equation*}
    I_{\mathrm{tot}} = I_{\mathrm{bulk}} + I_{\mathrm{surf}}
\end{equation*}

To accurately determine the bulk capacitance values, the~guard ring is also grounded during the \textit{C--V} measurements, as illustrated in Fig. \ref{fig:CV-setup}. The bulk capacitance is measured using a~testing signal with an amplitude of 2~V. The testing signal frequency of 1~kHz and 100~kHz was chosen for unirradiated and irradiated diodes, respectively. The measurements were performed by applying negative bias voltage\footnote{Absolute value of bias voltage is used for all discussions in the text of this paper.} on the backside of the diode. The tests were conducted at room temperature (RT) in a~probe station with monitored temperature and relative humidity (RH) for both unirradiated and irradiated diodes. All the irradiated diodes were tested before and after their annealing, performed for 80 minutes at 60~°C. 

The Top Transient Current Technique (Top-TCT) with red laser (660 nm) top illumination was used to verify $V_{\mathrm{FD}}$ and extract the electric field distribution and sign of space charge $N_{\mathrm{eff}}$ of the tested silicon diodes.


\subsection{Leakage current} \label{sec:current}

\begin{figure}[ht]
    \centering
    \includegraphics[width=10cm]{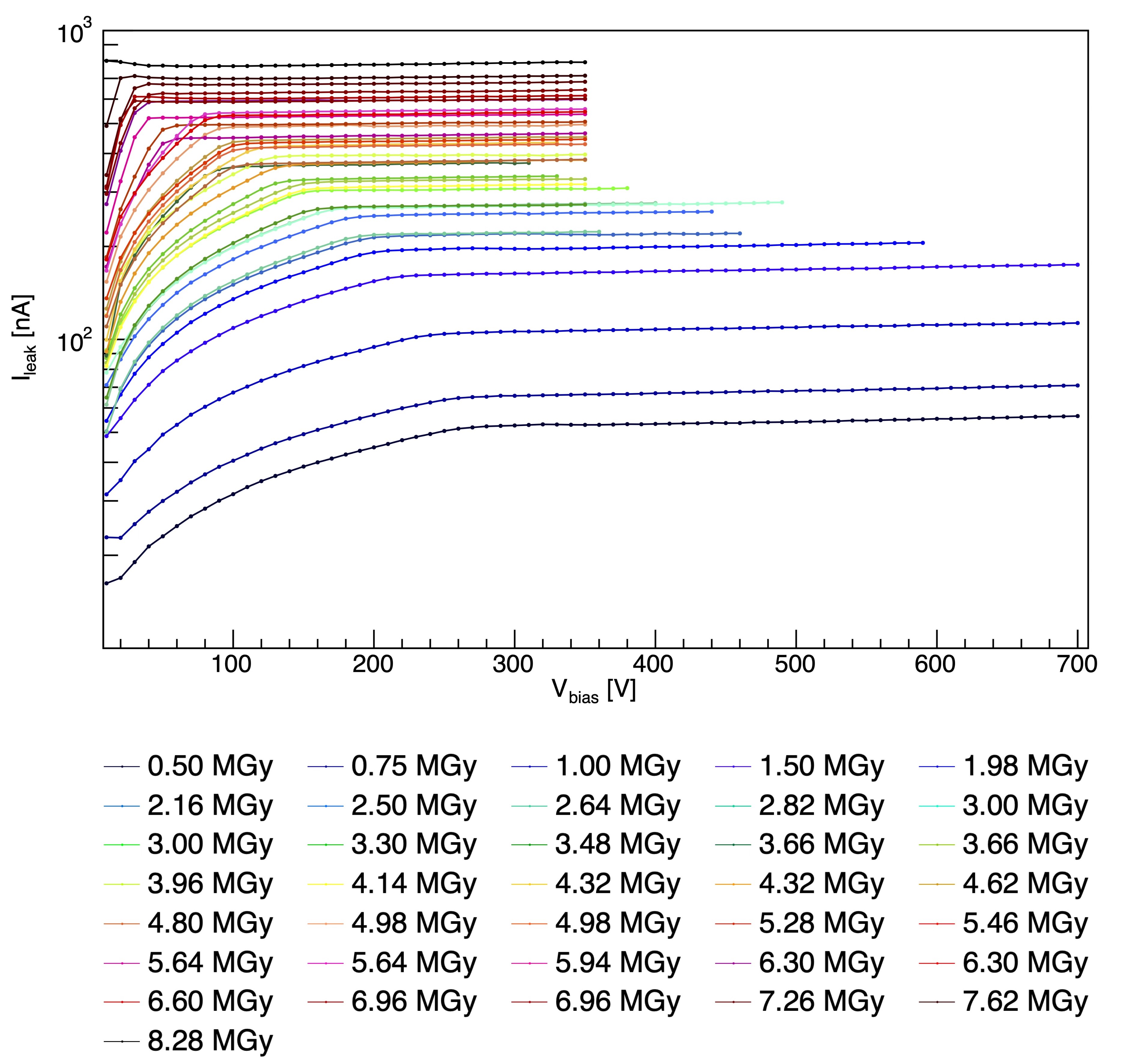}
    \caption{Bulk leakage current of diode B, irradiated to various TIDs ranging from 0.5 to 8.28 MGy, measured after annealing and plotted as a function of bias voltage.}
    \label{fig:IV-HPK}
\end{figure}

\textit{I--V} characteristics of the diode B, irradiated to various TIDs ranging from 0.5 MGy to 8.28 MGy, measured after annealing is shown in Fig. \ref{fig:IV-HPK}. The bulk leakage current increases with an increasing bias voltage until a full depletion of the diode is reached, then the current stabilizes. For fully depleted diodes, the bulk leakage current increases with an increasing TID.  The A and C diodes that are not shown in this plot exhibit similar behaviour.

\begin{figure}[h!]
\begin{subfigure}{0.49\textwidth}
\centering
\includegraphics[width=8cm]{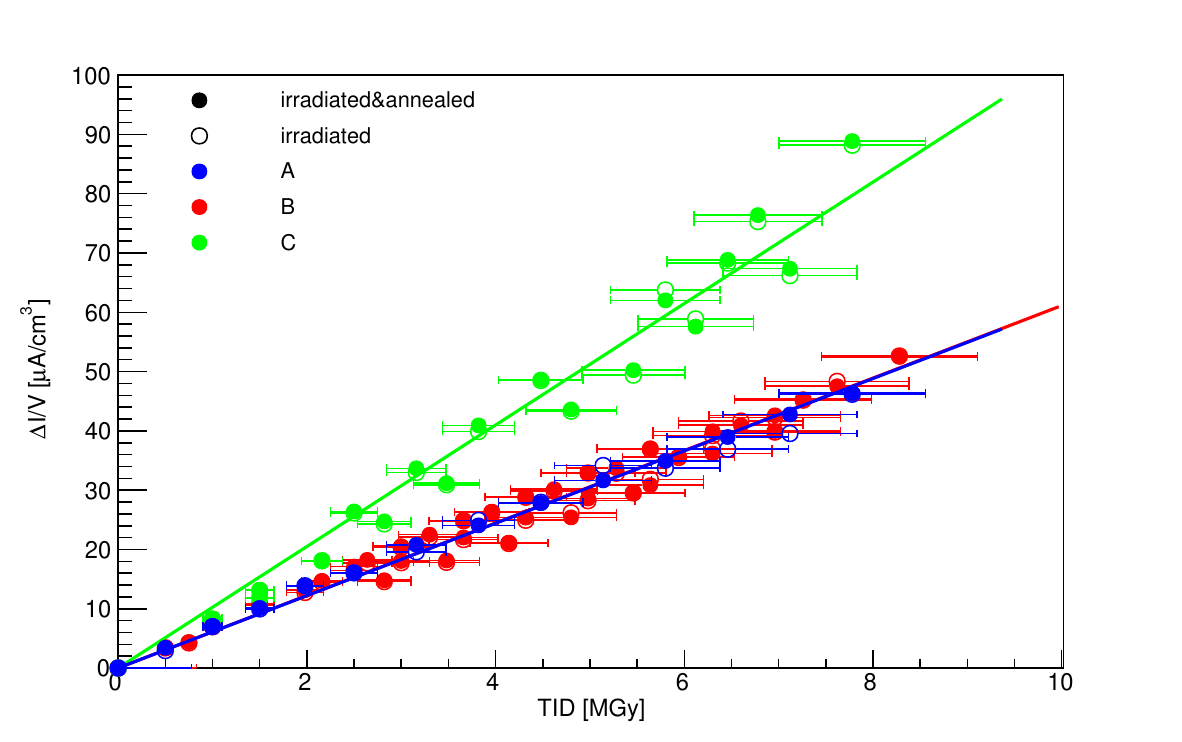}
\caption{}
\label{fig:TID-all}
\end{subfigure}
\begin{subfigure}{0.5\textwidth}
\centering
\includegraphics[width=8cm]{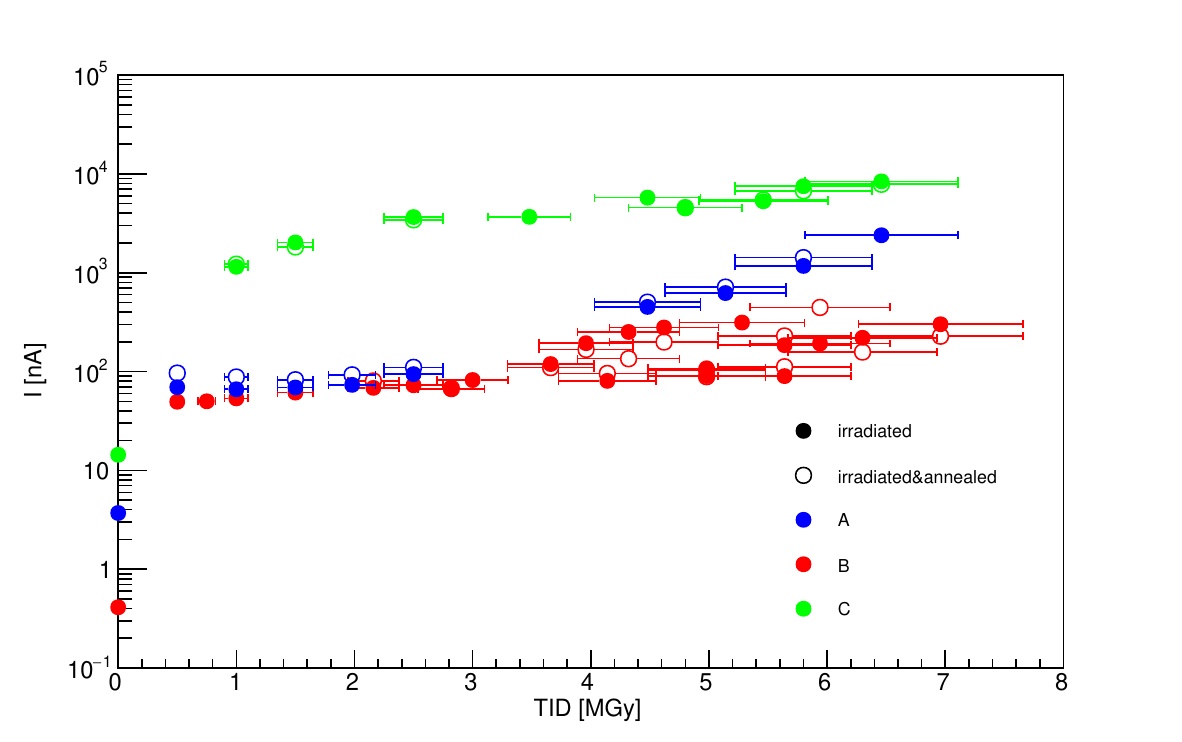}
\caption{}
\label{fig:IV-TID-surface}
\end{subfigure}
\caption{(a) Radiation-induced bulk leakage current measured per active volume $\Delta I/V$ of fully depleted diodes A, B, and C,  as a function of TID. Diodes are measured before and after annealing for 80 minutes at 60 °C. (b) Surface current of diodes A, B and C,  as a function of TID. All the measured currents are normalized to 20 °C.}
\label{fig:leakage-current}
\end{figure}

The radiation-induced bulk leakage current per active volume of the diode $\Delta I/V$ is plotted as a function of TID in Fig.~\ref{fig:TID-all}. The quantity $\Delta I$ is defined as a difference between the bulk leakage current of the fully depleted diode measured after and before irradiation, $V$ is the active volume contributing to the current. The leakage current is normalized to 20~°C and it is measured before and after the annealing (open and full marks in the plot). 
In the studied range of TIDs up to 8~MGy, the bulk leakage current in all types of $n^+$-in-$p$ silicon diodes shows a~linear increase with TID.

For the C diode, which has the highest bulk resistivity, i.e. the lowest boron concentration and highest oxygen concentration, an increase of the leakage current with a delivered TID is faster than for the A and B diodes, which have much lower resistivity. We can express the relation between $\Delta I/V$ and TID by the formula
\begin{equation*}
    \frac{\Delta I}{V}  = \alpha_{\mathrm{\gamma}} \cdot TID,
\end{equation*}

\noindent where $\alpha_{\mathrm{\gamma}}$ is the current related damage coefficient. The values of damage coefficient $\alpha_{\mathrm{\gamma}}$ obtained for each type of diode are listed in Table \ref{tab:damage}. The damage coefficients $\alpha_{\mathrm{\gamma}}$ measured for A and B diodes are in a very good agreement with the results obtained for gamma irradiated $p$-type diodes with similar initial resistivity of 4~k$\Omega \cdot$cm published in paper \cite{bib:liao} ($\alpha_{\gamma} \sim$ 5.1$\cdot10^{-6}$ A$\cdot$cm$^{-3}$ MGy$^{-1}$). 

A comparison of the bulk leakage current of gamma irradiated $n^+$-in-$p$ diodes measured before and after their annealing indicates that the bulk leakage current remains unchanged with the application of annealing. This observation again agrees with \cite{bib:liao}, where the measured leakage currents were stable for annealing temperatures up to 200~°C. Fig. \ref{fig:IV-TID-surface} shows the surface current of the studied diodes as a function of TID. The surface current values, which are  taken at $V_{\mathrm{FD}}$ +30~V, increase steeply for the initially delivered TIDs, then they become saturated. The surface current of diode A starts to increase again at high doses $>$~4 MGy. The surface current of diode C is one order of magnitude higher when compared to diodes A~and~B.

\begin{table}[ht]
    \centering
       \begin{tabular}{|c|c|c|}
    \hline
     Sample    & $\alpha_{\mathrm{\gamma}}$ [A$\cdot$cm$^{-3}$ MGy$^{-1}$] &   $\rho$ [k$\Omega \cdot$cm] \\ \hline
      A   & ($6.33\pm0.08$)$\cdot10^{-6}$   & $3.1\pm 0.1$\\ \hline
      B   & ($6.49\pm0.09$)$\cdot10^{-6}$     &$3.3\pm 0.1$  \\ \hline
      C   & ($10.20\pm0.30$)$\cdot10^{-6}$     & $24.0\pm4.0$ \\ \hline
    \end{tabular}
     \caption{Damage coefficient values $\alpha_{\mathrm{\gamma}}$ for each type of diode A, B, C and the~initial silicon resistivity $\rho$ \cite{bib:zatocilova}.}
    \label{tab:damage}
\end{table}

\begin{figure}[h!]
    \centering
    \includegraphics[width=10cm]{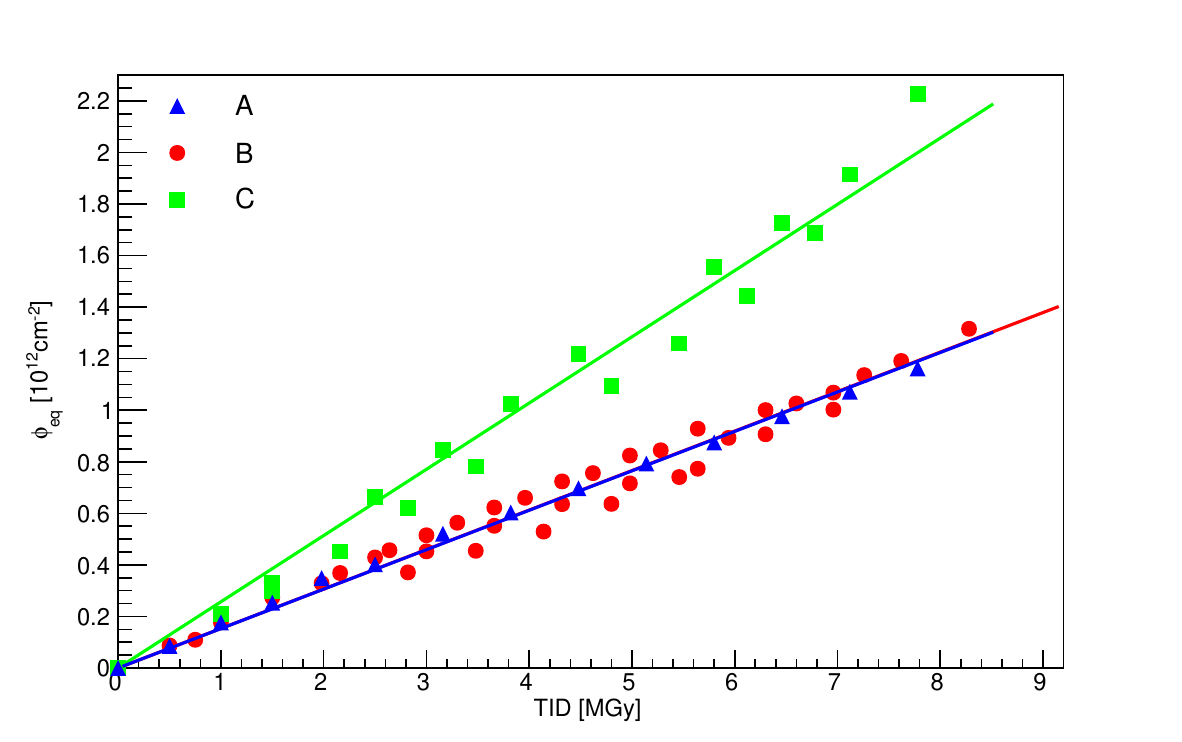}
    \caption{The relation between 1 MeV neutron equivalent fluence and TID delivered to studied diodes A, B, and C.}
    \label{fig:phi-eq}
\end{figure}

Assuming the linear increase of the radiation-induced leakage current with TID is caused by a displacement damage, we can estimate a conversion factor between TID in units of MGy delivered by gamma irradiation and 1~MeV neutron equivalent fluence $\phi_{\mathrm{eq}}$ in units of n$_{\mathrm{eq}}$/cm$^2$ as
\begin{equation*}
    TID\, [MGy] = k\cdot\phi_{\mathrm{eq}} [n_{\mathrm{eq}}/cm^2],
\end{equation*}

\noindent where $k$ is a proportionality constant, and $\phi_{\mathrm{eq}}$ is the fluence, that can be estimated by using the formula
\begin{equation*}
    \phi_{\mathrm{eq}} =\frac{\Delta I}{V}\frac{1}{\alpha}
\end{equation*}

\noindent with the damage parameter $\alpha$ = 3.99 $\times$ 10$^{-17}$ A/cm \cite{bib:moll}. The relation between the 1 MeV neutron equivalent fluence and TID is shown for all the studied types of diodes in Fig.~\ref{fig:phi-eq}. The estimated relation between the TID of 1 MGy and 1 MeV equivalent fluence $\phi_{\mathrm{eq}}$ is 1~MGy = 2.6 $\times$ 10$^{11}$ 1 MeV n$_{\mathrm{eq}}$/cm$^2$ for diode C ($\rho\approx$ 24 k$\Omega\cdot$cm), and  1 MGy = 1.6 $\times$ 10$^{11}$ 1 MeV n$_{\mathrm{eq}}$/cm$^2$ for diodes A (3.1 k$\Omega\cdot$cm) and B (3.3 k$\Omega\cdot$cm).

\begin{figure}[h!]
    \centering
    \begin{minipage}[t]{0.48\textwidth}
    \centering
    \includegraphics[width=\linewidth]{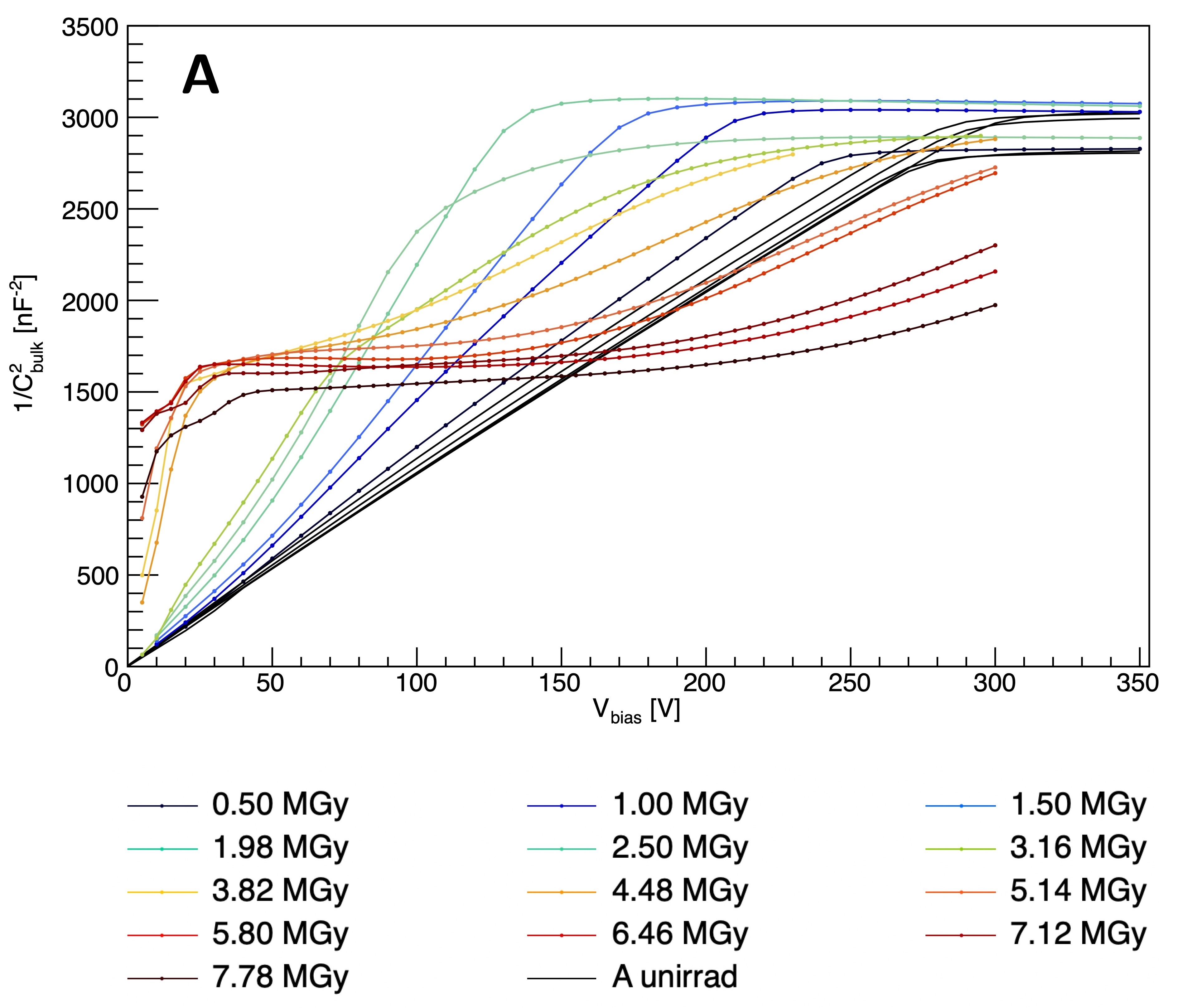}
\end{minipage}
\hfill
\begin{minipage}[t]{0.48\textwidth}
    \centering
    \includegraphics[width=\linewidth]{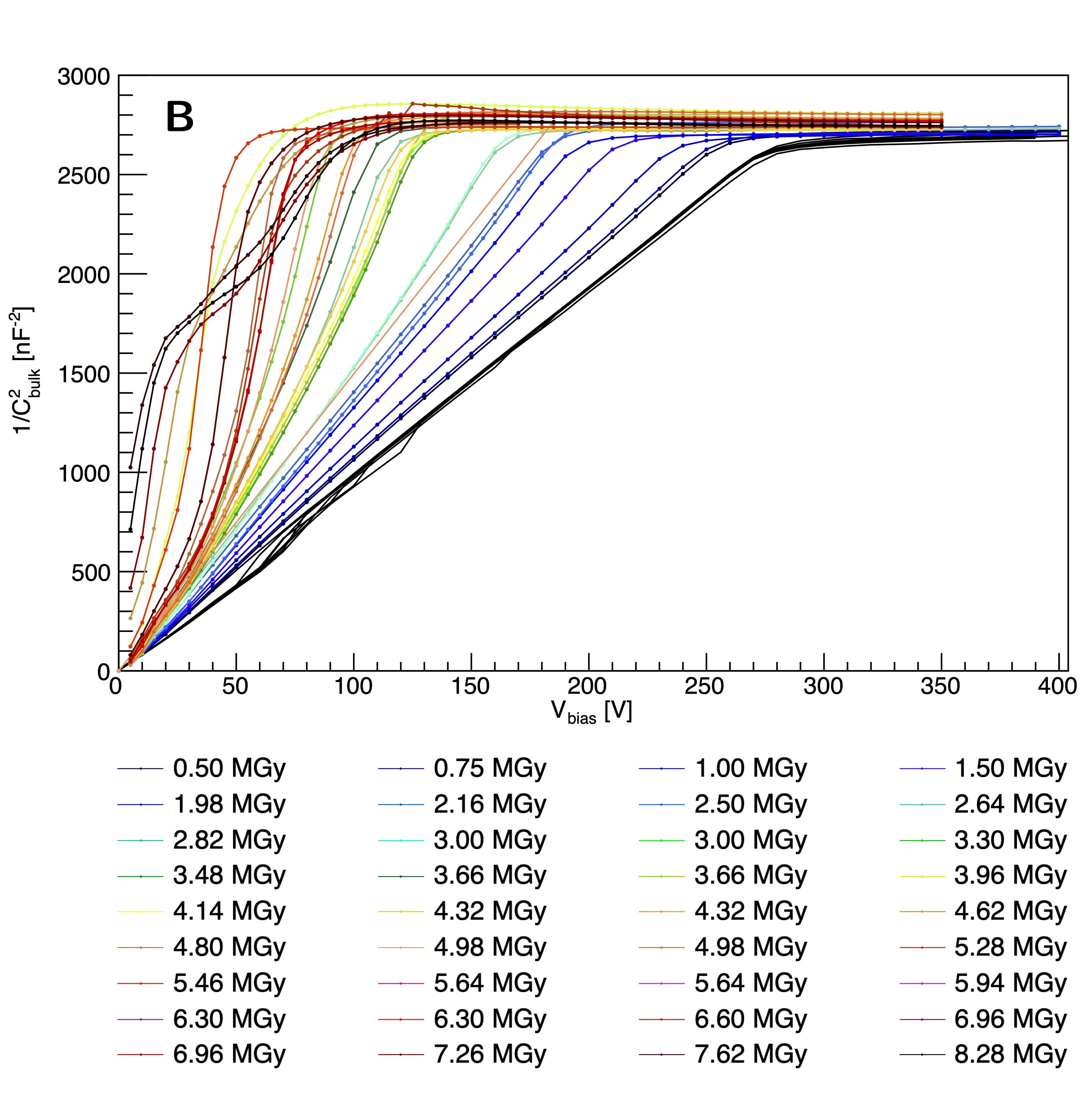}
\end{minipage}

\vspace{3mm}

\includegraphics[width=0.48\textwidth]{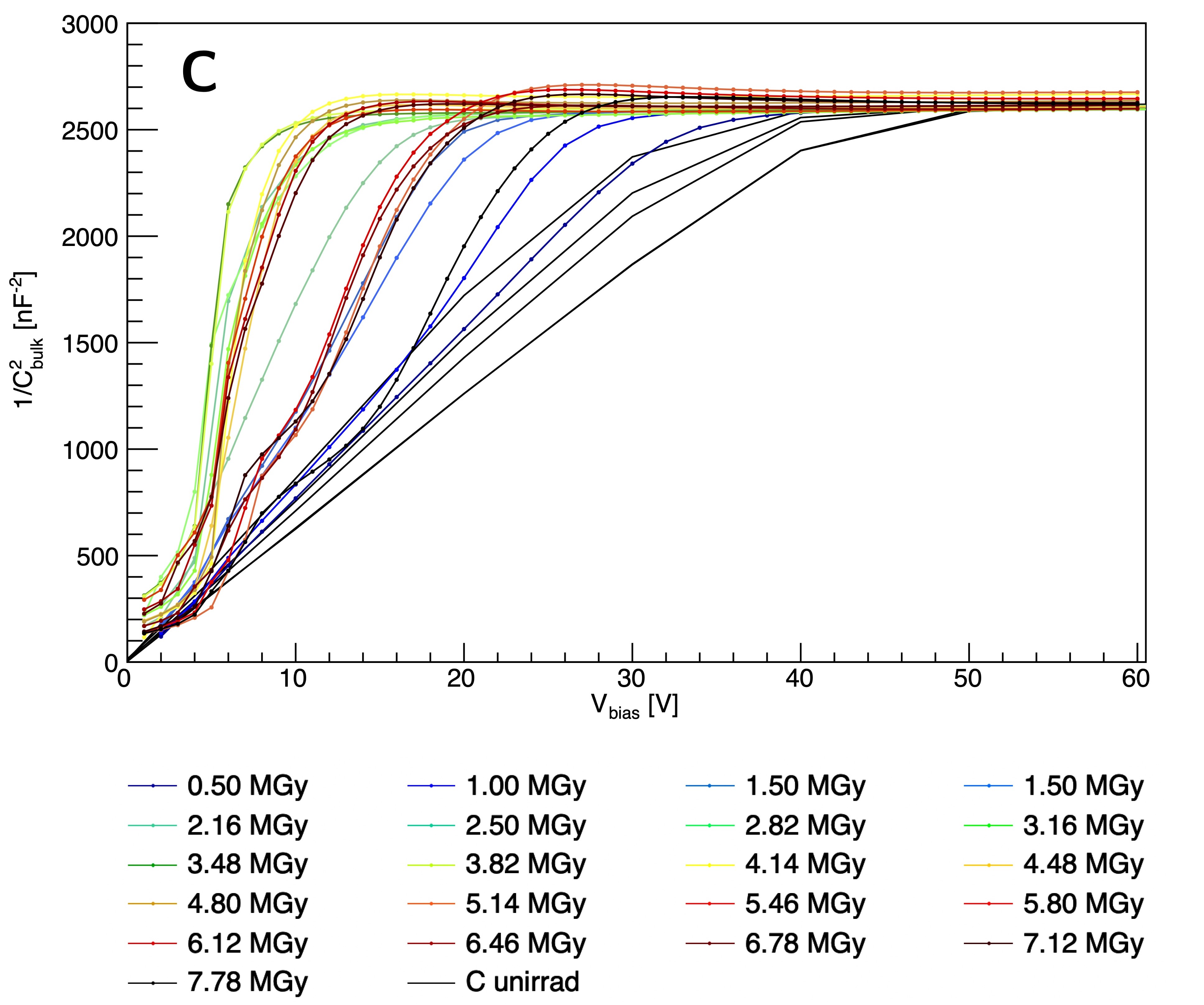}

\caption{\textit{C--V} characteristics of the diodes irradiated to various TIDs measured after annealing for 80 minutes at 60°C. Data obtained for unirradiated diodes (black lines) is shown for comparison.}
\label{fig:CV-characteristics}
\end{figure}
    \label{fig:CV-characteristics}

\subsection{\textit{C--V} measurement and full depletion voltage} \label{sec:CV-FDV}

The radiation-induced bulk defects result in modification of the effective space charge $N_{\mathrm{eff}}$ that manifests as an alteration in the full depletion voltage $V_{\mathrm{FD}}$ of a silicon diode. The $V_{\mathrm{FD}}$ is given as
\begin{equation*}
   V_{FD} = \frac{q|N_{eff}|\,d^2}{2\varepsilon},
\end{equation*}

\noindent where $d$ is an active thickness of the diode, $q$ is the elementary charge, and $\varepsilon$ is the permittivity of silicon. This equation assumes a constant space charge distribution over the volume of the diode. 

The~$V_{\mathrm{FD}}$ in this study was determined from the \textit{C--V} characteristics as the value of the bias voltage for which the linear increase of $1/C^2_{\mathrm{bulk}}$ dependence reaches its plateau. \textit{C--V} characteristics of diodes A, B, and C, irradiated to various TIDs, measured after their annealing are plotted in Fig. \ref{fig:CV-characteristics}. Plots also include equivalent dependence obtained for the unirradiated diode as a~reference. The evolution of $V_{\mathrm{FD}}$ with TID is clearly visible. However, determining the values of $V_{\mathrm{FD}}$ was not straightforward for highly irradiated diodes A and B due to the presence of two bumps in the $1/C^2_{bulk}$ dependence, with the plateau being reached only at the second bump. This is likely caused by the~creation of a double junction with high-field regions present on both diode sides, as observed previously in highly proton and neutron irradiated devices \cite{bib:casse}.  In this case, for bias voltages lower than $V_{\mathrm{FD}}$, the undepleted zone is located in the middle of the diode instead of at one of its sides. 

The values of $C_{\mathrm{bulk}}$ obtained for fully depleted unirradiated diodes correspond to 18.47, 19.30, and 19.52~pF for A, B, and C, respectively. These values are in a very good agreement (within 2 \%) with calculated values and prove a good quality of the measuring setup.

\begin{figure}[ht]
\begin{subfigure}{0.49\textwidth}
 \includegraphics[width=8cm]{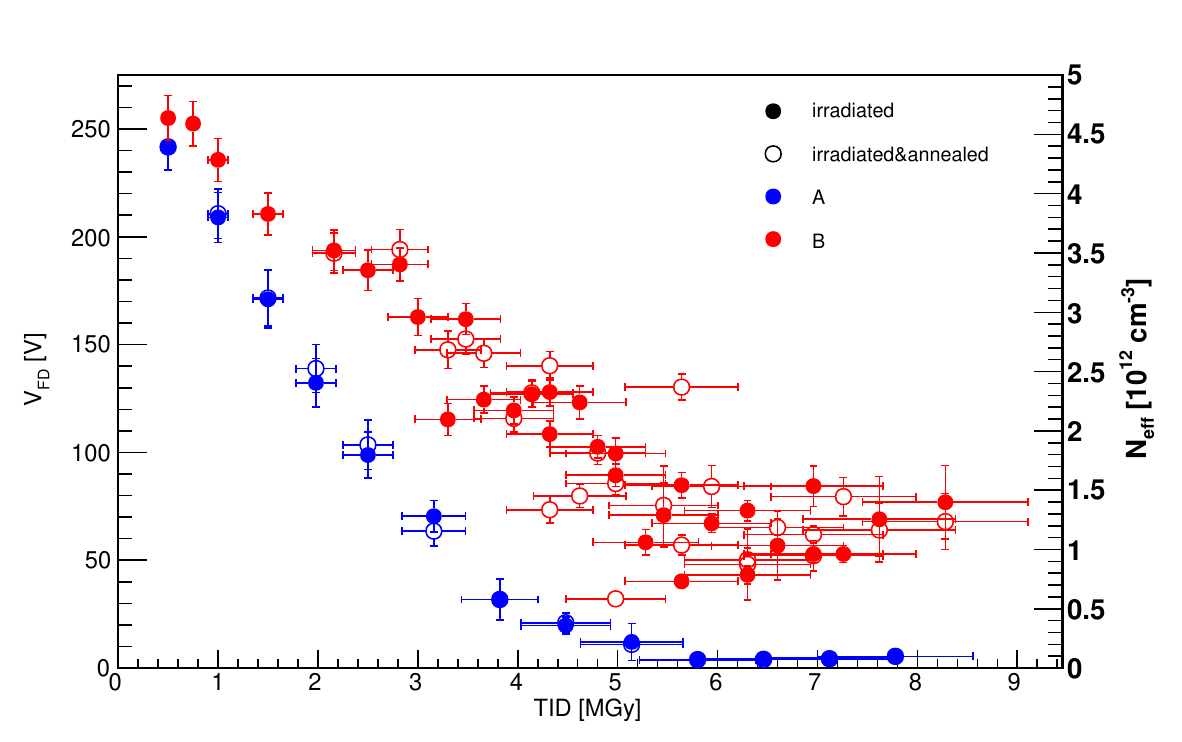} 
\caption{}
\label{fig:FDV-TID-AB}
\end{subfigure}
\begin{subfigure}{0.5\textwidth}
\includegraphics[width=8cm]{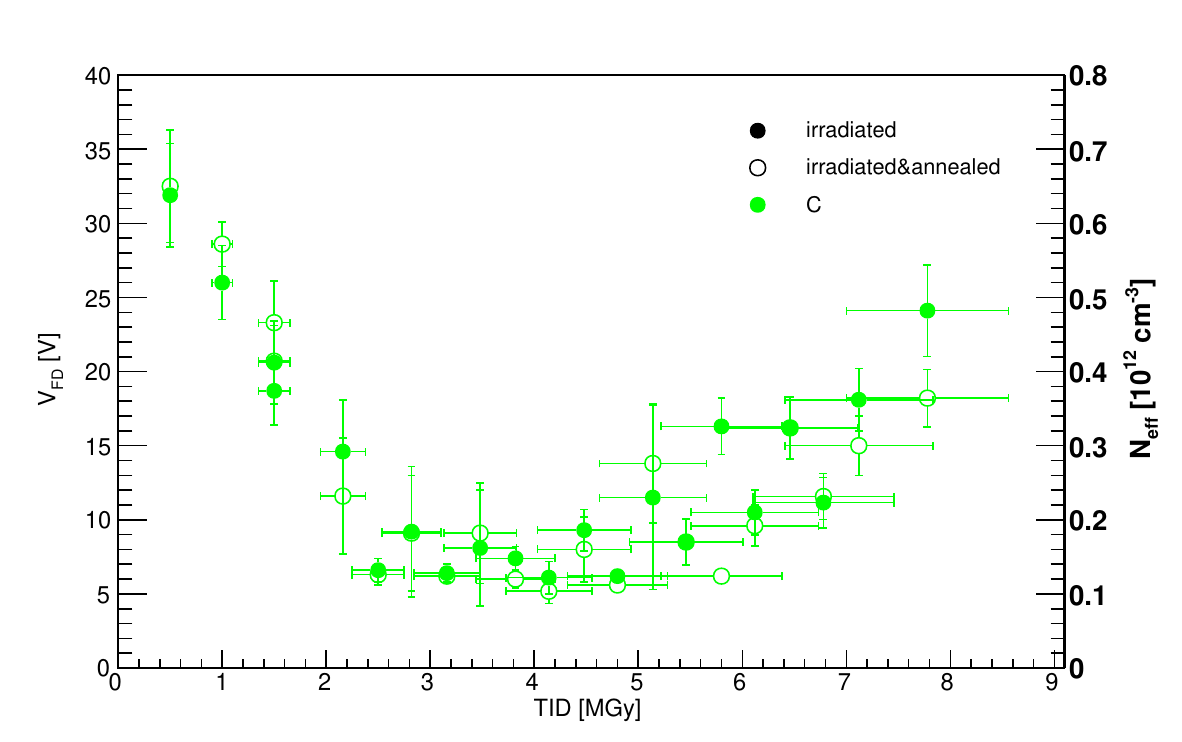}
\caption{}
\label{fig:FDV-TID-C}
\end{subfigure}
\caption{Dependence of the $V_{\mathrm{FD}}$ of irradiated A and B diodes (a) and C diode (b) on the delivered TID, measured both before (full marks) and after the standard annealing (open marks).}
\label{fig:FDV-TID}
\end{figure}

The~evolution of the full depletion voltage with an increasing TID is shown in Figs. \ref{fig:FDV-TID-AB} and \ref{fig:FDV-TID-C} for A, B, and C diodes, respectively. The~$V_{\mathrm{FD}}$, and thus also the $N_{\mathrm{eff}}$, of $p$-type silicon diodes measured both before and after their annealing significantly decreases with an increasing TID to a certain minimum value, and then increases again with an  increasing dose. The TID value, for which the $V_{\mathrm{FD}}$ reaches its minimum, reflects the initial resistivity of the diode. For the C diodes, having the highest initial resistivity of $\approx$ 24 k$\Omega\cdot$cm, the minimum value of $V_{\mathrm{FD}}$ is reached for the TID of 4.14 MGy. The A and B diodes, which have a comparable initial resistivity of $\approx$ 3.1 - 3.3 k$\Omega\cdot$cm, reach the minimal $V_{\mathrm{FD}}$ at TID close to 6 MGy. For the diodes of type A the measured $V_{\mathrm{FD}}$ is not increasing in the measured interval of TIDs and for TIDs higher than 6 MGy stays at a value $\sim$10 - 20 V. However, it should be noted that the $V_{\mathrm{FD}}$ values for diodes A were determined from the first bump in the \textit{C-V} characteristics, not at bias voltages where the value of $C_{\mathrm{bulk}}$ reaches the value of geometric capacitance.  The results obtained by the TCT method with the A diode irradiated to the highest dose of 7.78 MGy show, that the diode is fully depleted at $\sim$70 V, as discussed in section \ref{sec:TCT}. The fact that the geometric capacitance is not reached for this diode is not fully understood and needs further study. It is most likely related to a complex space charge distribution (i.e., visible double peak in TCT voltage scan of induced current pulses) in the diode \cite{bib:casse, bib:Li, bib:eremin, bib:verbitskaya, bib:kramberger}.

The initial decrease in $N_{eff}$ is in accord with findings published in Ref.~\cite{bib:liao}, where $p$-type diodes were subjected to gamma irradiation up to 2 MGy. The 30\% decrease in $N_{eff}$ measured for the delivered TID of $\sim$2 MGy in~\cite{bib:liao} is very similar to our observations, as shown in Fig.~\ref{fig:FDV-TID} for sample B at the equivalent TID. Due to the lower maximal TID reached, no saturation or increase in $N_{eff}$ was observed in~\cite{bib:liao}.

No systematic effect of annealing on the obtained $V_{FD}$ was observed both in our study, as well as in Ref.~\cite{bib:liao}, where isochronal annealing - with a 10 °C step and 15 minutes annealing at each temperature - was conducted up to the maximal temperature of 150 °C. The reaction of gamma irradiated diodes on annealing is significantly different from the annealing effect observed for proton irradiated $p$-type samples, where a strong beneficial annealing manifested by decrease in $N_{eff}$ is observed, followed by long-term annealing with an increase in $N_{eff}$ \cite{bib:segneri}.

\subsection{TCT studies of gamma irradiated $n^+$-in-$p$ silicon diodes} \label{sec:TCT}

The transient-current technique (TCT) with laser pulses is a well established method for studies of silicon detectors \cite{bib:eremin2, bib:kramberger2}.  The principle of the TCT is the observation of transient currents induced on readout electrodes by short pulses of a laser light with a wideband current amplifier. The measurements were made with the system produced by Particulars \cite{bib:web}. In this work, pulses of red laser light ($\lambda$ = 640 nm), shorter than 1~ns, were directed to the top surface ($n^+$) of the DUT. Because of the short penetration depth of red light in silicon charge carriers are released very close to the surface. In a reverse biased $n^+$-in-$p$-$p^+$  diode, electrons generated close to the $n^+$ surface are immediately collected by the electrode while holes drift through the depleted region because of the electric field. As follows from the Ramo’s theorem, the induced current is proportional to the velocity of carriers.

TCT measurements were conducted only on diodes A and~B, as diodes of type C lack a necessary opening in metallization. Diodes irradiated to the lowest and highest TIDs were measured.

\begin{figure}[h!]
\centering

\begin{subfigure}[t]{0.49\textwidth}
    \centering
    \includegraphics[width=\linewidth]{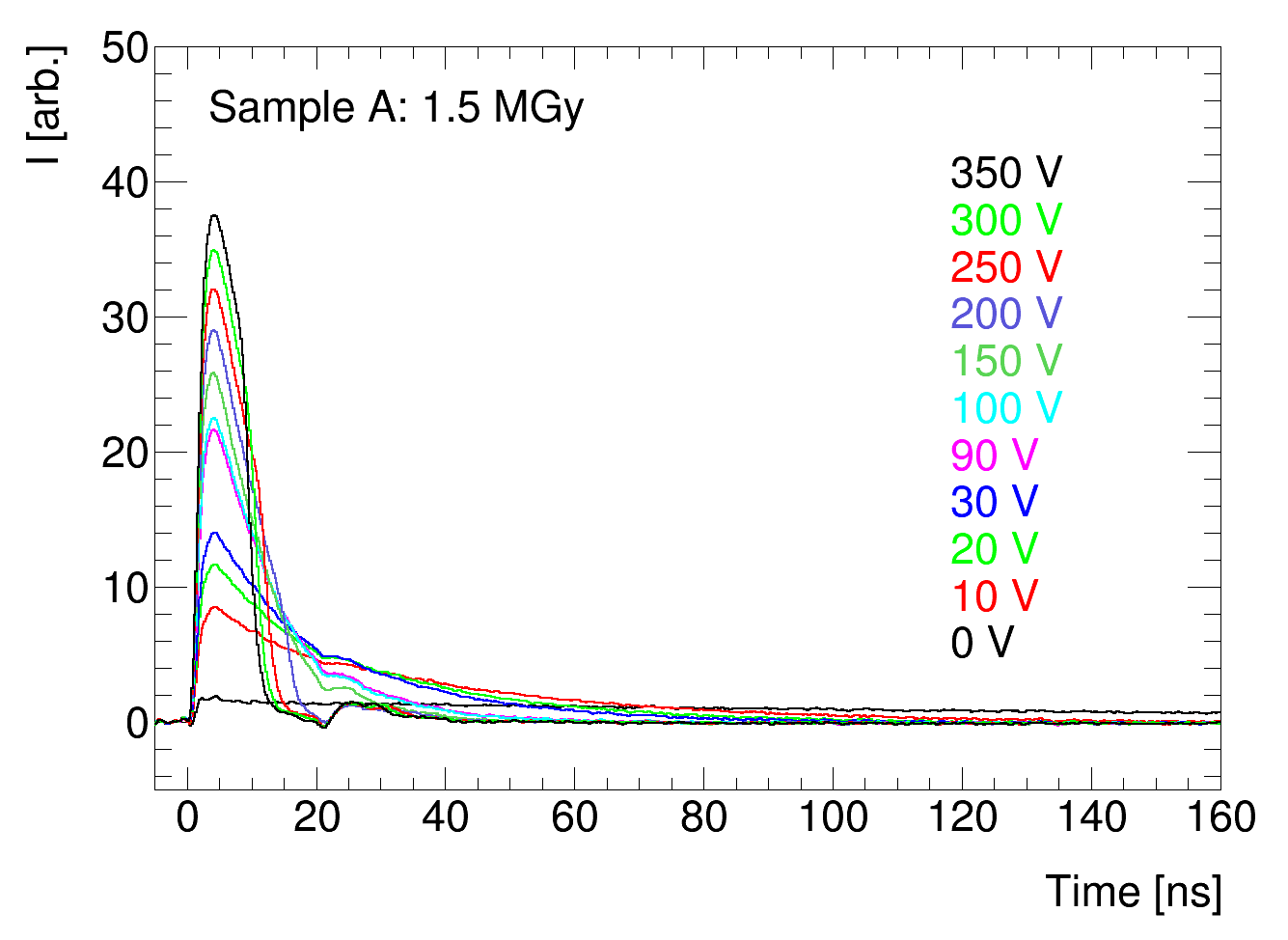}
    \caption{}
\end{subfigure}
\hfill
\begin{subfigure}[t]{0.49\textwidth}
    \centering
    \includegraphics[width=\linewidth]{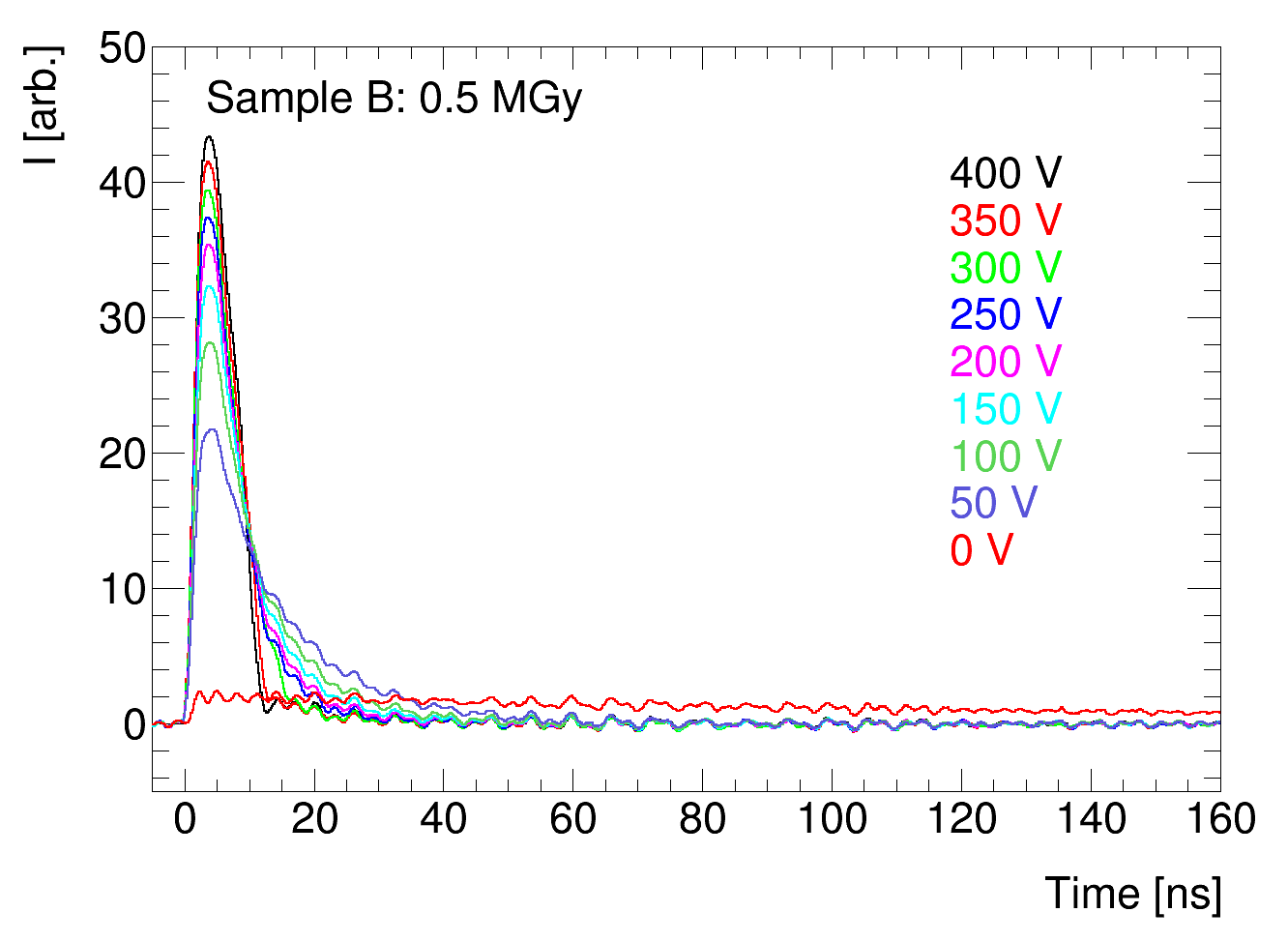}
    \caption{}
\end{subfigure}

\caption{Top-TCT results for low-irradiated diodes A (a) and B (b). Diodes A and B were irradiated to 1.5 and 0.5 MGy, respectively.}
\label{fig:low-dose}
\end{figure}

Fig.~\ref{fig:low-dose} shows the induced current pulses from a TCT voltage scan between 0~and 400~V performed for low-irradiated diodes A and B. Diode A was irradiated to 1.5~MGy, while the diode B to 0.5~MGy. In both cases, the high electric field on the $n^+$ side reflects in the high induced current immediately after a laser pulse at $t=0$. The induced current is decreasing with time as the carriers drift towards the $p^+$ side. This confirms that at these low TIDs, the bulk of the diodes stays $p$-type, i.e. the space charge sign inversion (SCSI) is not observed and the diodes still exhibit the $n^+$-in-$p$ structure. $V_{FD}$ can be estimated from the shapes of the current pulses in Fig.~\ref{fig:low-dose}. As long as  the diode is fully depleted, the end of the current pulse has a visible edge. At bias below $V_{FD}$ the induced current pulse approaches exponentially the bias line as the carriers drift towards the not depleted part of the diode. The $V_{FD}$ obtained from the TCT measurement is between 150~and 200~V for diode A and between 250~and~300~V for diode B.

These results are in very good agreement with the values of $V_{FD}$ obtained from the \textit{C–V} characteristics, where the $V_{FD}$ is 170~V for diode A and 255~V for diode B.

\begin{figure}[h!]
\centering

\begin{subfigure}[t]{0.49\textwidth}
    \centering
    \includegraphics[width=\linewidth]{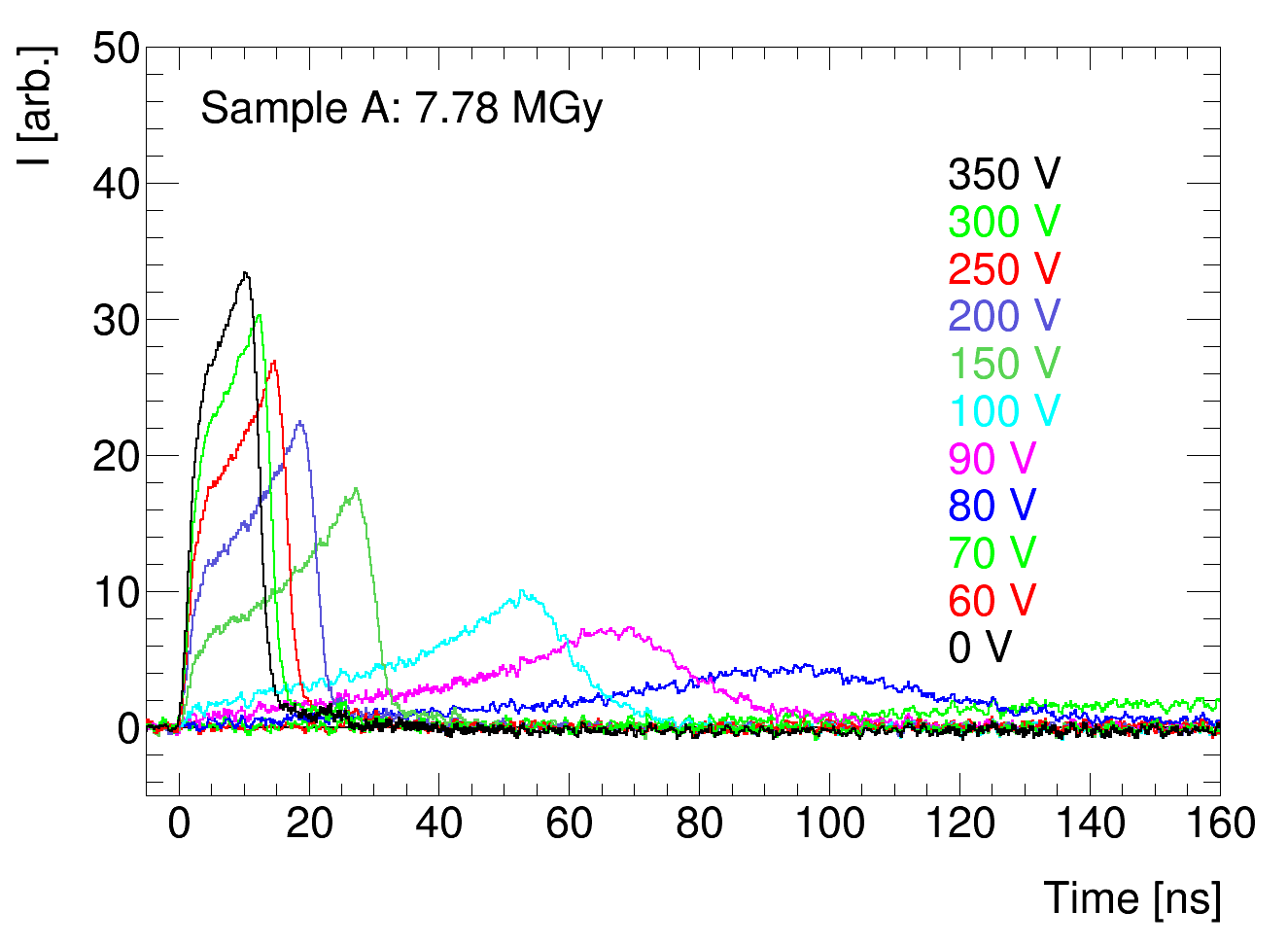}
    \caption{}
\end{subfigure}
\hfill
\begin{subfigure}[t]{0.49\textwidth}
    \centering
    \includegraphics[width=\linewidth]{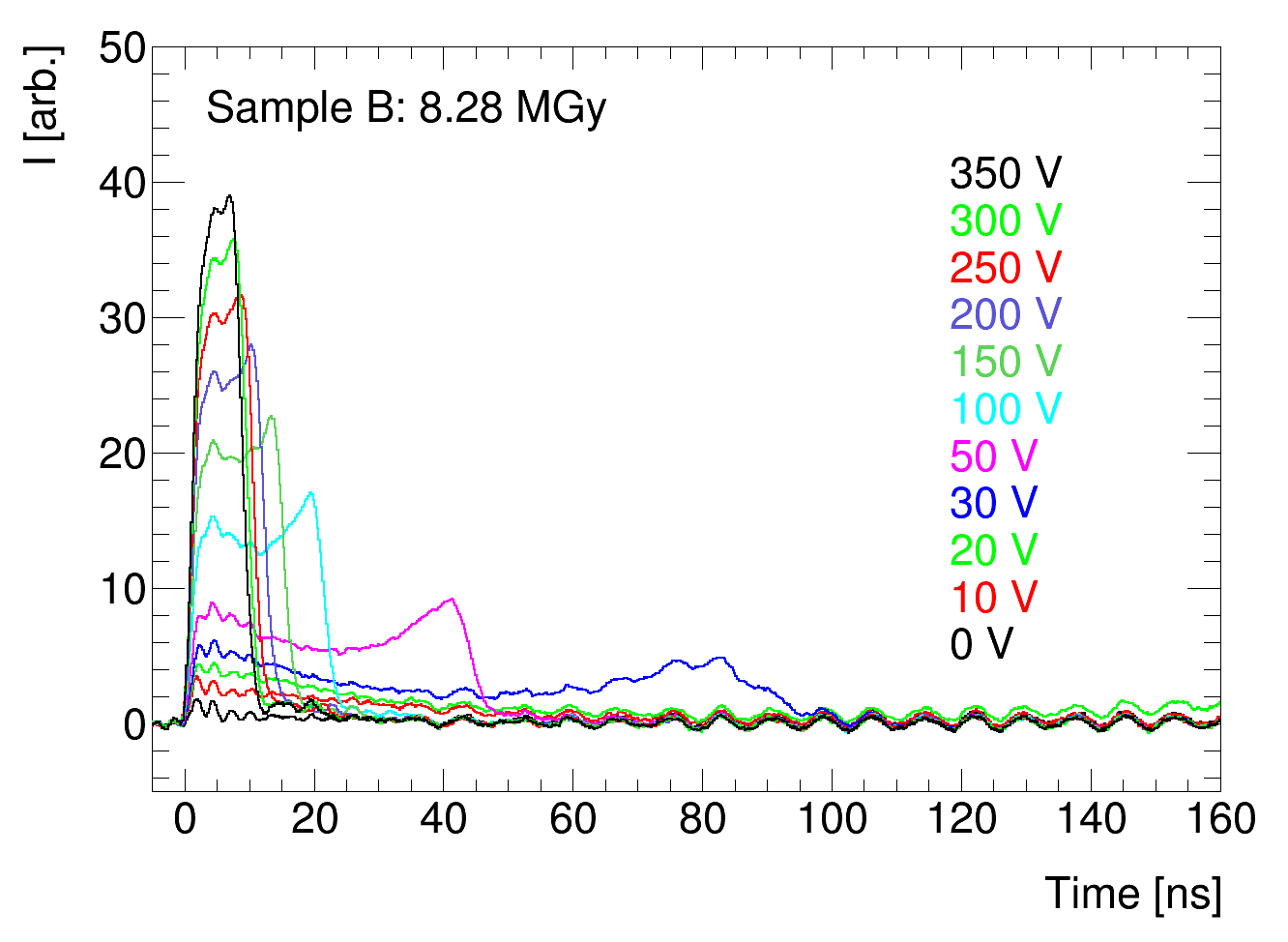}
    \caption{}
\end{subfigure}

\caption{Top-TCT results for high-irradiated diode A (a) and diode B (b). Diode A was irradiated with 7.78 MGy, and diode B with 8.28 MGy.}
\label{fig:high-dose}
\end{figure}

The Top-TCT voltage scans for bias voltages ranging from 0~to 350~V performed on highly irradiated diodes A (7.78~MGy) and B (8.28~MGy) are presented in Fig.~\ref{fig:high-dose}. The pulse shapes for both diodes indicate that the $pn$-junction is located on the back side of the diodes. It means that the high dose of gamma irradiation has inverted the bulk from $p$-type to $n$-type. In the case of the highly irradiated diode~A, the~Top-TCT voltage scan reveals no signal up to a bias voltage of 60~V, indicating no depletion on the $n^+$ side. However, at a bias voltage of 70~V, a~very long pulse is observed, suggesting the $V_{FD}$ around 70~V. The induced current is low immediately after a laser pulse because the carriers are released in a low electric field on the $n^+$ side of the diode. As time progresses, the signal increases because the holes drift in the increasing electric field as they approach the~$p^+$ side. This clearly indicates an $n$-type bulk device with $pn$-junction on the back side.

\begin{figure}[h!]
    \centering
    \includegraphics[width=7cm]{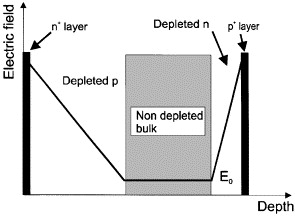}
    \caption{Qualitative electric field distribution in an irradiated silicon with double junction \cite{bib:Li, bib:eremin, bib:verbitskaya, bib:kramberger}.}
    \label{fig:el-field-distr}
\end{figure}

For diode~B, the Top-TCT voltage scan indicates that there is a~signal already at a~bias voltage of $\sim$~30~V, which increases slightly at longer times. This suggests that the diode is fully depleted between 30 and~40~V. However, the pulse exhibits two peaks, indicating a $p$-type bulk on the $n^+$ side and an~$n$-type bulk on the $p^+$ side, with a~neutral bulk region in between. Nonetheless, the electric field is present throughout. At higher bias voltages, the~pulse is nearly flat, slightly higher towards longer times, indicating an $n$-type bulk but with a low space charge concentration. The~shape of the pulses indicates the double junction with high field regions on both diode sides. The~qualitative electric field distribution of the double junction is shown in Fig.\ref{fig:el-field-distr}.


\section{Conclusions} \label{sec:conclusions}

The radiation damage of high resistivity $p$-type silicon diodes caused by high doses of gamma irradiation from $^{60}$Co source, reaching the TIDs up to 8.28 MGy, has been studied. The study was carried out on three types of $n^+$-in-$p$ diodes with different initial resistivities of 3.1, 3.3, and 24.0 k$\Omega\cdot$cm, but similar diode geometries. The main goal of the study was to characterize the evolution of the $V_{FD}$ with TID by measuring \textit{C--V} characteristics and the gamma-radiation induced displacement damage by measuring \textit{I--V} characteristics of the studied diodes. It was observed that the bulk leakage current increases linearly with TID, and the damage coefficient depends on the initial resistivity of the silicon diode. The damage coefficients of the diodes with initial silicon resistivities of $\approx$ 24, 3.3, and 3.1 k$\Omega\cdot$cm were determined to be 10.20, 6.49, and 6.33 $\times$ 10$^{-6}$ A$\cdot$cm$^{-3}\cdot$MGy$^{-1}$, respectively. Assuming the linear increase of the radiation-induced leakage current with TID is caused by a displacement damage, we estimated the relation between the TID of 1~MGy and the 1~MeV neutron equivalent fluence $\phi_{\mathrm{eq}}$ as follows: for the diode with $\rho$~=~24 k$\Omega\cdot$cm, $1\;\mathrm{MGy} = 2.6\times10^{11}\;1\,\mathrm{MeV}\;\mathrm{n_{eq}/cm^2}$, and for the diodes with $\rho$ = 3.3 and 3.1 k$\Omega\cdot$cm, 1~MGy = 1.6 $\times$ 10$^{11}\;1\,\mathrm{MeV}\;\mathrm{n_{eq}/cm^2}$. The surface current increases significantly  for the initial delivered TID, then it saturates.

The effective doping concentration significantly decreases with increasing TID, before it starts to increase at a specific TID value. The initial decrease of the $V_{\mathrm{FD}}$ could be explained by a deactivation of electrically active boron dopants. The high doses of gamma irradiation invert the bulk from $p$-type to $n$-type, as confirmed by the applied Top-TCT technique.  A similar effect was observed in $p$-type silicon samples irradiated by hadrons \cite{bib:moll2}. Our study reveals that a diode with higher initial resistivity, i.e. with lower or compensated boron doping, reaches the minimum value of $V_{\mathrm{FD}}$  at  lower TID compared to diodes with a lower initial resistivity.

Another important conclusion of this study is that annealing for 80 minutes at 60~°C has no effect on the gamma radiation-induced damage in $p$-type silicon. There are no changes in the $V_{\mathrm{FD}}$  or bulk leakage current values observed after such annealing. This annealing behavior of gamma-irradiated $n^+$-in-$p$ silicon diodes is quite different from that observed for hadron-irradiated $n^+$-in-$p$ silicon diodes. The fact that neither the $V_{\mathrm{FD}}$ nor the bulk leakage current changes with annealing could originate from predominantly immobile defects.

\section*{Acknowledgments} \label{sec:aknowledgements}
This work was supported by the Ministry of Education, Youth and Sports of the Czech Republic coming from the projects LM2023040 CERN-CZ and LTT17018 Inter-Excellence and  FORTE - CZ.02.01.01/00/22\_008/0004632. Part of this work was also funded by the DPG within the framework of GRK 2044/2.


\end{document}